# Conversion of multilayer graphene into continuous ultrathin $sp^3$-bonded carbon films on metal surfaces

Dorj Odkhuu[1], Dongbin Shin[2], Rodney S. Ruoff[3], and Noejung Park[1,2★]

[1]Interdisciplinary School of Green Energy and Low Dimensional Carbon Materials Center,

[2]Department of Physics, Ulsan National Institute of Science and Technology (UNIST), Ulsan 689–798, Korea

[3]Department of Mechanical Engineering and the Materials Science and Engineering Program, The University of Texas, Austin, Texas 78712

★noejung@unist.ac.kr

**The conversion of multilayer graphenes into $sp^3$-bonded carbon films on metal surfaces (through hydrogenation or fluorination of the outer surface of the top graphene layer) is indicated through first-principles computations.** ***The main driving force for this conversion is the hybridization between carbon $sp^3$ orbitals and metal surface $d_z^2$ orbitals.*** **The induced electronic gap states in the carbon layers are confined in a region within 0.5 nm of the metal surface. Whether the conversion occurs depend on the fraction of hydrogenated (fluorinated) C atoms and on the number of stacked graphene layers. In the analysis of the Eliashberg spectral functions for the $sp^3$ carbon films on diamagnetic metals, the strong covalent metal-$sp^3$ carbon bonds induce soft phonon modes that predominantly contribute to large electron-phonon couplings, suggesting the possibility of phonon-mediated superconductivity. Our results suggest a route to experimental realization of large-area ultrathin $sp^3$-bonded carbon films on metal surfaces.**

Diamond is known for its extraordinary thermal conductivity and mechanical hardness, and the production of large-area crystals has been a long term goal. The conventional synthetic methods employed for the production of artificial diamond usually require extremely high pressure and temperature[1,2], or plasma-enhanced chemical vapor deposition (CVD)[3–5]. Advances in the growth of large-area graphene[6,7] and bilayer graphene through multilayer graphenes[8–10] suggest another possibility: the chemical conversion of a few layers of graphene into $sp^3$-bonded carbon films. For example, previous theoretical results showed that the complete hydrogenation[11] or fluorination[12] of a single layer of graphene can yield a thermodynamically stable $sp^3$-bonded carbon layer. Indeed, partial hydrogenation of monolayer graphene[13] and evidence of complete fluorination of monolayer graphene membranes[14] have been experimentally achieved. Similarly, the conversion of a few layers of graphene has attracted, and hydrogenation of the outer surfaces of bilayer graphene[15,16] to achieve all $sp^3$-bonded carbon films has been treated theoretically. We suggest however that such an approach will be challenging experimentally as extensive hydrogenation or fluorination was stated to be needed on both the top and bottom surfaces.

Here we present a promising alternative, namely the conversion of graphene layers on a metal surface into $sp^3$ carbon films through surface hydrogenation or fluorination. From the practical point of view, the commonly used methods for growing graphene films on a metal substrate, either through CVD or as a result of the transfer of grown graphenes, are already providing appropriate target systems[6–10,17] for the study of the $sp^2$ to $sp^3$ conversion of carbon films on metal surfaces. Our computation results suggest a method of preparing an ultrathin 'diamond' layer over even very large area by conversion of appropriate multilayer graphene on metal surfaces. The resulting structure could also be viewed as a carbon-based metal-insulator junction that may reveal novel two-dimensional (2D) phenomena, as we suggest in this study, or as a new type of electronic material[18].

We first investigated the effect of hydrogenation or fluorination on the free surface of the AB-stacked (Bernal-type stacking) bilayer graphene. Simulating experimental generation of gas phase atoms[13,19], hydrogen or fluorine atoms were considered to be available on the surface. Figure 1a is a schematic of the bilayer graphene before the adsorption of such gas atoms (denoted by I). Experimental studies reported that the hydrogen chemisorption coverage of a single graphene side is less than half of the carbon atoms[19]. It was however suggested that bilayer or multilayer graphene can have more extensive hydrogenation, which can be up to half the coverage of the outer surfaces[15,20]. In Fig.1, our super-cell model contains hydrogen atoms that could cover half the carbon atoms on the outer surface. After the chemisorption of hydrogen atoms, the two graphene layers can be separated by the ‘van der Waals’ (vdW) distance, as shown in Fig. 1b, or can form interlayer C–C bonds, as shown in Fig. 1c. The overall energetics is described in Fig. 1d.

The half coverage hydrogenation can exist in various configurations. Top views of the patterns named as ‘boat-like’, ‘zigzag-like’, and ‘chair-like’[11,12] are presented in the insets of Fig. 1d. In the boat-like configuration, two hydrogen atoms are located at the two ends of the bridge of the C–C bond. In the zigzag-like configuration, each carbon atom in every second zigzag chain is bonded to a hydrogen adatom. In the chair-like configuration, the C–H bonds are equally spaced. These three configurations, in which the carbon layers are separated by the vdW distance, are together categorized as configuration II in Fig. 1. The chair-like configurations induce $sp^3$ hybridization in the carbon atoms, leading to $sp^3$ dangling bonds in the inner side of the top graphene layer. This drives formation of covalent C–C bonds between the upper and lower graphene layers, as shown in Fig. 1c (denoted as III). In contrast, dangling bonds are essentially not present for the boat-like and zigzag-like configurations owing to the pairing of two neighboring p orbitals in the form of local π bonding[21]. Thus, the boat-like and zigzag-like configurations are much more stable than the chair-like one either in the configuration II or III.

*When the graphene bilayer is positioned on transition metal surfaces, the energetics for conversion to $sp^3$ films are significantly different than for the isolated bilayer*. In Fig. 2, we consider the same adsorption-induced conversion of bilayer graphene on (0001) surface of Co (a few other metals are also considered, as discussed below). Besides the existence of Co below the graphene layers, the adsorption configurations of C–H bonds presented in Figs. 2a, b, and c, and the insets of Fig. 2d (configurations I, II, and III) are the same as those shown in Fig. 1. The favored adsorption site of carbon atoms is found to be atop Co atoms of the interface, as shown

in Fig. 2a, which agrees with previous studies[22,23]. The energy differences between the boat-like, zigzag-like, and chair-like configurations in group II are nearly the same as those presented in Fig. 1. *However, it is noteworthy that the presence of Co greatly stabilizes configuration III; this is in obvious contrast to the situation without the metal substrate.* This stabilization occurs because of the saturation of the otherwise unstable $sp^3$ dangling bonds with Co surface states. Further, as shown in Fig. 2d, the transition barrier from configuration II to configuration III is negligibly small (about 0.5 kcal $mol^{-1}$).

Bond lengths and bond angles in configuration III reveal obvious $sp^3$ carbon bond features. The calculated interlayer C–C and intralayer C–C bond lengths are 1.58 Å and 1.52 Å, respectively, and these values are comparable to the known $sp^3$ bond length (1.54 Å) in bulk diamond. The tetrahedral angles of the C–C–C bond within the layer and that formed with the interlayer C–C bond are 110.2° and 108.5°, respectively; these values are close to the tetrahedral bond angle of 109.5°. Fluorination of the graphene bilayer on Co(0001) revealed similar features as those observed for the hydrogenated cases. Details are available in Supplementary Table S1.

To understand the underlying electronic origin of the $sp^2$ to $sp^3$ conversion, we investigated the electronic structures of the converted $sp^3$ carbon films on Co(0001). The projected electronic band structure and the projected density of states (PDOS) of carbon atoms in the outermost layer are presented in Fig. 3a. Pristine single-layer graphene has the cone-like energy bands touching the Fermi level at the K point of the 2D Brillouin zone. Such characteristics, originated from the delocalized π electronic states, largely disappeared upon the conversion from $sp^2$ to $sp^3$ bonding. The top of the valence bands, in which the $p_x$ and $p_y$ orbital states are degenerate, and the bottom of the conduction band, which is characteristic of non-degenerate $p_z$ orbital state, are set far apart from each other, resulting in a wide direct gap of about 3.18 eV at the Γ point. DFT calculations yielded similar wide band gaps for graphane[11,12] and diamane[15,24] (which are $sp^3$ carbon structures with full hydrogenation of top and bottom surfaces of graphene and AB-stacked bilayer graphene). The minority spin states, presented in Fig. 3b, reveal the same features; thus, there is no measurable magnetization in the outermost layer.

Figure 3c shows the band structure and PDOS in the majority spin state of the carbon layer at the interface with Co. The corresponding results for the minority spin states are shown in Fig. 3d. The in-plane p bands depicted in green ($p_x$) and orange ($p_y$) are almost identical to the corresponding projected bands of the outer layer (Fig. 3a and b). However, the hybridization features between the carbon $p_z$ and cobalt $d_z^2$ orbitals are prominent throughout the energy level, leading to the drastic changes in energy and dispersion of $p_z$ bands. For comparison, the projected electronic structures of majority and minority spins in the interface Co atoms are presented in Figs. 3e and f, respectively. Such strong hybridization causes spin exchange-splitting in the carbon layer. The magnetic moment in the first carbon layer is calculated to be about –0.08 $\mu_B$ per carbon atom and anti-parallel to the Co spin direction, which is twice the value observed in experiments for graphene adsorbed onto the Co(0001) surface[25]. These metallic states penetrate into a thicker $sp^3$ carbon layers, constituting the metal-induced gap state (MIGS).

To estimate the range of the MIGS, we considered the possibility of converting graphene multilayers of various thickness (see further discussion on thickness-dependent energetics below) to $sp^3$ carbon films; for example, we considered the case of five graphene layers stacked in the AB configuration on Co(0001), as shown in Supplementary Fig. S1. The PDOS resolved into each layer (Supplementary Fig. S1c) shows that the first carbon layer (L1) at the interface with Co has electronic states persisting over the Fermi level with large exchange-splitting. The evanescent tails of such gap states decay rapidly and almost disappear in the third layer (L3). This indicates that such $sp^3$ C films with a thickness of only approximately 0.5 nm from metal surface can likely provide good electric isolation.

We next investigated the effect of the extent of hydrogen coverage on the transformation of AB-stacked graphene bilayer into an $sp^3$ carbon film. Figure 4a shows the formation energy and interlayer spacing (in the minimum energy configuration) between the two graphene layers on Co(0001) as a function of the hydrogen coverage of the outer (i.e., top) surface. The formation energy drops monotonically with increasing coverage and reaches its minimum at the maximal coverage ($C_2H$). (Note that the more negative formation energy indicates the structure is more stable). Previously, a similar trend was also found in diamane structures[15]. When the coverage of the outer surface is less than one quarter of the maximal coverage ($C_8H$), the two carbon layers are separated by 3.12 Å, that is essentially the interlayer vdW distance. For H coverage greater than half of the maximal coverage ($C_4H$), the formation of the interlayer C–C covalent bonds is favored and the carbon atoms on both layers turn to have predominantly the $sp^3$ characteristics. This dependence on the coverage is largely consistent with Angus and Hayman's formulation[26] for the compositions of $sp^3$ and $sp^2$ bonds in the hydrogen-treated diamond-like hydrocarbons, as discussed in detail in Supplementary information.

We now report the effect of thickness on the stability of fully hydrogenated ($C_2H$) $sp^3$ carbon films on metal substrates. We consider two different stacking configurations of graphene layers, namely AB and AA. The AB-stacked graphene was converted into a cubic diamond-like layer, whereas the AA-stacked graphene transformed into a hexagonal diamond-like layer. The side views of the cubic and hexagonal diamond-like overlayers are presented in the insets of Fig. 4b: the cubic diamond consists of three alternating layers (denoted by α, β, and γ) and the hexagonal diamond consists of two alternating layers (denoted by α and β). Figure 4b shows the formation energies of cubic and hexagonal diamond-like ultrathin films on the Co(0001) substrate. The negative formation energies indicate that the $sp^3$-bonded diamond-like structures on the Co surface are thermodynamically and structurally stable for thicknesses of up to eight carbon layers.

The preferable conversion energetics from $sp^2$ layers into $sp^3$ carbon films can be applied to the cases of other metal surfaces. The reactivity of metal surfaces can be roughly classified into two groups, depending on the strength of their interaction with graphene: the 'physisorption group' (e.g., Cu, Pt, Al, Ag, or Au), corresponding to the surfaces on which graphene weakly physisorbs through vdW attraction, and the 'chemisorption group' (e.g., Ni, Co, or Pd), corresponding to the surfaces on which graphene develops relatively strong chemical bonds through a hybridization between its p orbitals and the metal d orbitals[22]. In the present study, we selected Co(0001), Ni(111), and Cu(111) because of the convenience associated with the periodic super-cell calculation: the lattice mismatches of these metals with the graphene are only 1.6, 1.2, and 3.6%, respectively. As expected, the conversion energetics for graphene layers on Ni(111) revealed features that were almost similar to those of Co(0001).

A more remarkable point is the stabilization of the configuration III on the Cu(111) surface. As discussed in ref. 22, the overlayer of pure graphene (configuration I) and the one-side hydrogenated graphenes (configurations II) physisorb onto Cu(111) through the vdW interaction. However, for configuration III, the $sp^3$ dangling bonds at the interface develop covalent bonds with surface Cu atoms, similar to observations for Co and Ni. The calculated relative energies of the hydrogenated bilayer graphene on Ni(111) and Cu(111) substrates (along with the case of Co(0001) and the case without metal substrate) are summarized in Table 1. Clearly, *the stabilization by hybridization between the $sp^3$ dangling orbital and the $d_z^2$ metallic orbital in* *configuration III applies not only to the chemisorbing metal surfaces, such as Co and Ni, but also to the relatively less reactive metals, such as Cu.*

The electronic structures of the $sp^3$-bonded carbon layers on metal surfaces, shown in previous paragraphs, revealed similar features as those of hole-doped diamond. This motivated us to study the features of electron-phonon coupling that may lead to phonon-mediated superconductivity. The superconductivity of hole-doped diamond (the substitution of B for C) has attracted broad interest[27,28], and a possible modulation of critical temperature ($T_c$) in a single $sp^3$ carbon layer (i.e., graphane) through B doping has also been studied recently[29]. To investigate the electron-phonon coupling characteristics in the metal/diamond interface structures, we selected a two-layer $sp^3$ carbon structure on Cu(111) that possesses a perfect diamagnetic electronic structure. The phonon DOS (PHDOS) of the hydrogenated $sp^3$-bonded carbon layers on Cu(111) is shown in Fig. 5a. For comparison, the PHDOS of the two-layer $sp^3$ structure hydrogenated on both sides (i.e., diamane) is also presented in Fig. 5a. Both systems show quite similar PHDOS characteristics in the frequency modes ranging from 400 to 1300 $cm^{-1}$; the overall patterns are analogous to those of pristine diamond[30]. The sharp peaks around 1200–1300 $cm^{-1}$ can mainly be attributed to the shear motion between the carbon and hydrogen layer in the outer surface (Details are given in Supplementary Fig. S2). The C–C in-plane and out-of-plane vibrations dominate the medium-frequency region, contributing to a few peaks at around 600, 800, and 1000 $cm^{-1}$. For the cases of B-doped graphane and diamonds, the softening of these carbons stretching modes with increasing doping concentration has been discussed in relation to the electron-phonon coupling, thus increasing $T_c$ [29,31]. By comparison with the diamane, a PHDOS characteristic of the $sp^3$ layer on Cu is the presence of soft modes in the frequency range below 250 $cm^{-1}$ that are due to the vibration of Cu–C bonds: in-plane modes with a peak at about 140 $cm^{-1}$ and out-of-plane modes with a peak at about 230 $cm^{-1}$ (indicated by arrows).

Figure 5b shows the Eliashberg's spectral function $\alpha^2F(\omega)$ and the electron-phonon coupling $\lambda(\omega)$ calculated for the $sp^3$-bonded carbon layers on Cu(111). The overall PHDOS characteristics dominates the $\alpha^2F(\omega)$ throughout the range of phonon frequencies. The calculated $\lambda(\omega)$ increases in two steps at low and medium frequencies, at which the aforementioned Cu–C and C–C modes exert a critical effect. More specifically, $\alpha^2F(\omega)$ possesses a sizeable spectrum (indicated by an arrow) at the low frequency region, that extends up to 250 $cm^{-1}$, which can be attributed to soft phonons of the metal/$sp^3$ carbon interface, which contribute considerably to the integrated $\lambda$ of 0.25. The high-frequency peaks in $\alpha^2F(\omega)$ at around 1200–1300 $cm^{-1}$, originated from the shear motion of the outermost C and H layers, provides a negligible contribution to $\lambda$[29], because $\lambda$ is a function of the inverse of $\omega$, i.e., $\lambda=2\int \alpha^2F(\omega)\,\omega^{-1}d\omega$.

As a complementary study, in order to compare with the B-doped diamond systems, we also considered the effect of hole-doping on $\lambda$. As is summarized in Supplementary Table S2, the quantity $\lambda$ increases with increasing doping concentration and reaches 0.81 at the doping concentration of 12.5%, which is substantially greater than the value of 0.56 obtained for a 10% B-doped diamond[31]. This sharp increase in $\lambda$ is mainly attributed to the enhancement of the electronic DOS at the Fermi level ($N(E_F)$). *These results imply that the $\lambda$ value can be engineered by choosing the metal substrate that can promote $N(E_F)$ to the interface $sp^3$ carbon layers.* To demonstrate this conjecture, we repeated our calculations for the $sp^3$-bonded carbon layers on the (111) surface of Pt: the relatively large $N(E_F)$ indeed leads to a large $\lambda$ of 0.45 *without doping effect.* Details are given in Supplementary Table S2.

The computation for a thicker $sp^3$ carbon layer on metal surface is too much demanding, but we conjecture that the results of electron-phonon calculations presented above should be valid for thicker diamond layers because the σ-bonded Fermi level states in the $sp^3$ framework are rather localized in the vicinity of the metal/diamond interface. To verify the accuracy of the present computation, we calculated the electron-phonon features of known materials (i.e., $MgB_2$). Details are given in Supplementary Table S2. For the case of the $sp^2$ graphene overlayer on Cu(111), $\lambda$ is quite small, and thus the possibility of phonon-mediated superconductivity is negligible. Even though the calculated $\lambda$ and $T_c$ of our $sp^3$ carbon layers on metals are lower than those of $MgB_2$, our results indicate that metal/diamond interfaces can serve as another candidate for superconductivity study.

The scaled preparation of multilayer graphene on metal substrates and our computational results suggest a route to preparation of very large area ultrathin $sp^3$-bonded carbon films that would represent an entirely new material with the potential for a very wide range of applications.

**Methods**

The DFT calculations were performed using the projector augmented wave pseudo-potential method[32], as implemented in the Vienna *ab initio* simulation package[33]. Exchange and correlation interactions between electrons were described with the generalized gradient approximation formulated by Perdew, Burke, and Ernzerhof (PBE)[34]. The long-range dispersion corrections for the interlayer interaction (Figs.1,2,3) were taken into account within the semi-empirical DFT-D2 approach suggested by Grimme[35]. Spin polarization was taken into account for all the calculations. The PHDOS, Eliashberg spectral function $\alpha^2F(\omega)$, and electron-phonon coupling $\lambda(\omega)$ were calculated using the linear response theory within the density functional perturbation theory[36], as implemented in the Quantum ESPRESSO[37]. The critical temperature $T_c$ was estimated using McMillan's formula for the Eliashberg equation[38],

$$T_c = \frac{\omega_{ln}}{1.2} \exp\left[ \frac{-1.04(1+\lambda)}{\lambda(1-0.62\mu^*)-\mu^*} \right]$$

where $\omega_{ln}$ is the logarithmic average frequency and $\mu^*$ is the screened Coulomb pseudo-potential. The values of $T_c$ shown in Supplementary Table S2 were calculated with the typical choice of $\mu^*=0.1$ [31]. Some more details for supercell configurations, relaxation scheme, and *k*-point samplings are given in Supporting Information.

**Acknowledgements**

This research was supported by the Basic Science Research Program through the National Research Foundation of Korea (NRF), funded by the Ministry of Education (NRF-2013R1A1A2007910). RSR appreciates support from his Cockrell Family Endowed Regents Chair.

**Author contributions**

D.O. performed the calculations and prepared the manuscript. D.S. carried out the electron-phonon coupling calculations. R.S.R contributed scientific ideas and guidance including on the contents/preparation of the manuscript. N.P. designed research and is responsible for the calculations and also in guiding the content of the manuscript. All authors have given approval to the final version of the manuscript.

**Additional information**

Supplementary information is available in the online version of the paper. Reprints and permissions information is available online at www.nature.com/reprints.

**Competing financial interests**

The authors declare no competing financial interests.

**Figure 1 | The energetics related to the adsorption-induced geometric changes of bilayer graphene. a–c,** The geometry before (**a**) and after (**b**) the hydrogen adsorption, and after formation of interlayer C–C bonds (**c**). **d,** Relative energies for each configuration and energy barrier. The insets in **d** show three different hydrogen adsorption configurations onto the outer

surface of graphene: boat-like, zigzag-like, and chair-like. The larger gray and the smaller blue balls represent the carbon and hydrogen atoms, respectively.

**Figure 2 | The conversion energetics of bilayer graphene into $sp^3$ carbon film on metal surface. a–d,** The same energetics and geometric changes as the ones that were shown in Fig. 1 but with Co(0001) underneath the graphene bilayer. Large pink balls are the cobalt atoms.

**Figure 3 | Electronic structures of the two-layer $sp^3$ carbon film on Co(0001). a,b,** Majority (**a**) and minority (**b**) spin band structures and PDOS of carbon atoms on the top layer. **c,d,** The same majority (**c**) and minority (**d**) states projected onto carbon atoms at the interface. **e,f,** The same majority (**e**) and minority (**f**) states projected onto the top Co layer at the interface. The Fermi level is set to zero energy.

**Figure 4 | The dependence of stability of $sp^3$ carbon films on the adsorption coverage and thickness. a,** The hydrogen coverage-dependent formation energy (filled square) and optimized C–C interlayer distance (open circle) of bilayer graphene on Co(0001). The vertical line represents the critical coverage ($x \approx 1/3$) beyond which the interlayer C–C bonds are favored. The insets show the optimized structures corresponding to the coverage just below (x=0.25) and above (x=0.5) the critical coverage. **b,** The thickness-dependent formation energies of the fully hydrogenated ($C_2H$ at the outer surface) $sp^3$ carbon films on Co(0001). Filled squares and open circles refer to the formation of cubic and hexagonal diamond configuration, respectively. The insets show the side view of the optimized seven-layer diamond films on Co. The carbon atoms

in different stacking layers (αβγ for cubic and αβ for hexagonal diamond configurations) are differentiated with gray scale. Other atomic symbols follow the same convention used in Fig. 2.

**Figure 5 | Phonon and electron-phonon coupling features of the $sp^3$ carbon/Cu(111) interface. a,b,** Phonon density of state (**a**) and Eliashberg function $\alpha^2F(\omega)$ and the electron-phonon coupling constant $\lambda(\omega)$ (**b**) for the two-layer $sp^3$ carbon film on Cu(111). In **a**, the phonon density of state for diamane is shown with green line for comparison. The arrows indicate the C–Cu vibration modes.

**Table 1 | Relative energies of configurations I, II, and III depicted in Figs. 2a, b, and c on Co(0001), Ni(111), and Cu(111). In each case with different metal, the energy reference is the configuration I. The energies are given in kcal $mol^{-1}$ of hydrogen adatoms.**

| Groups | Configurations | Pristine | Cobalt | Nickel | Copper |
|---|---|---|---|---|---|
| I | | 0.0 | 0.0 | 0.0 | 0.0 |
| | boat-like | –36.4 | –37.3 | –37.1 | –38.1 |
| II | zigzag-like | –35.4 | –36.5 | –36.2 | –37.2 |
| | chair-like | –15.6 | –16.2 | –16.4 | –18.8 |
| III | | –11.3 | –45.4 | –47.8 | –45.2 |

**Figure 1:**

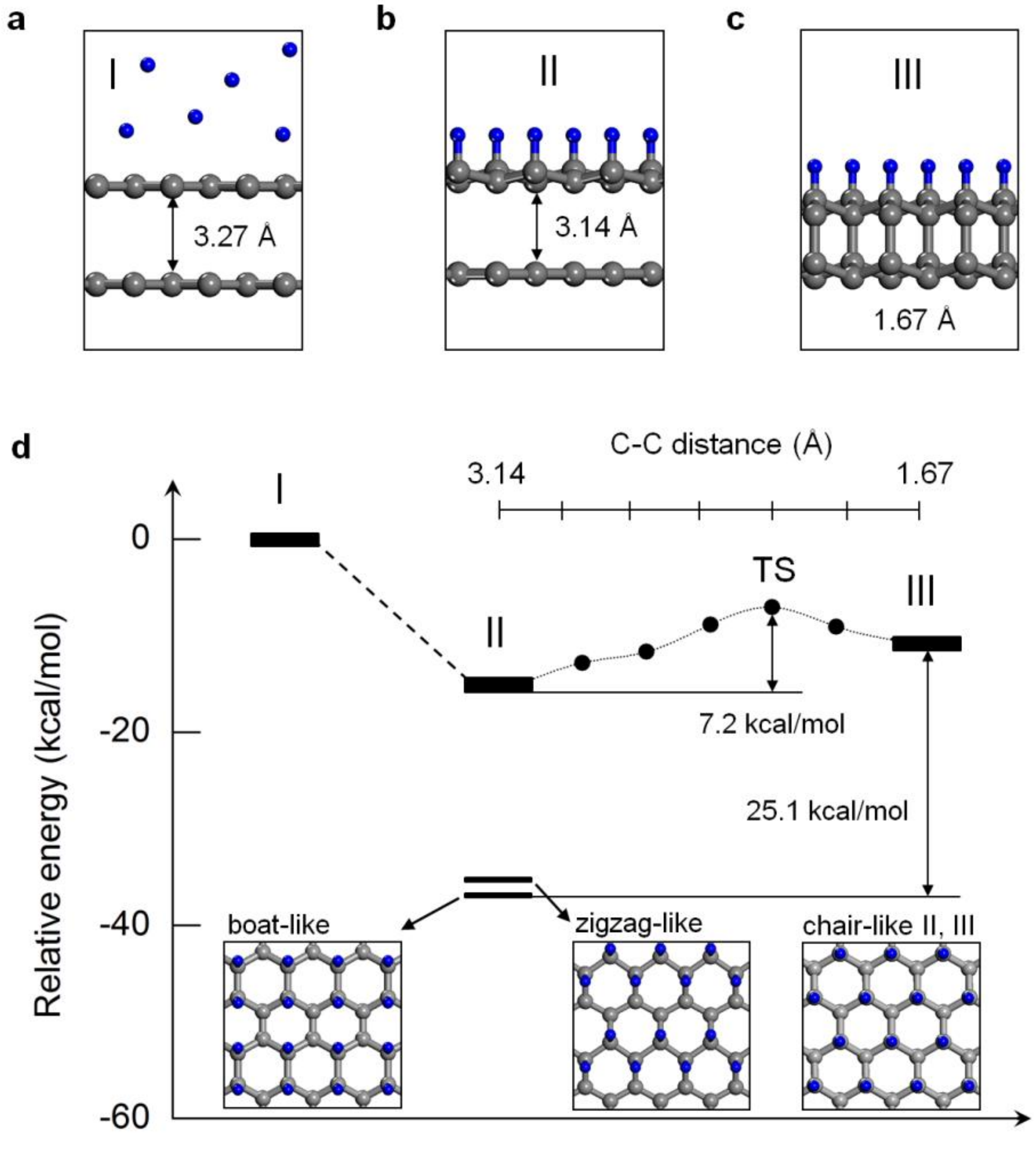

**Figure 2:**

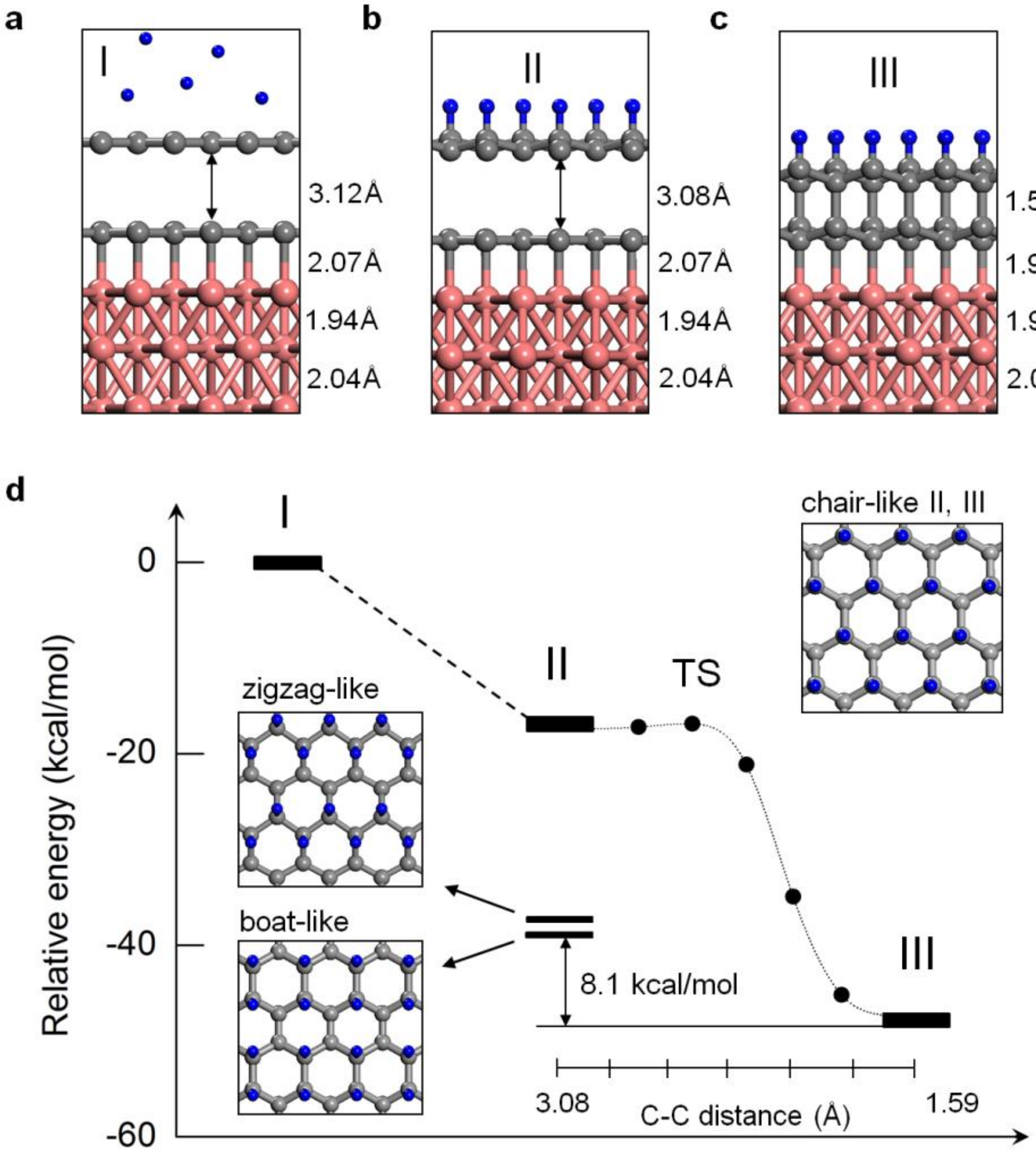

**Figure 3:**

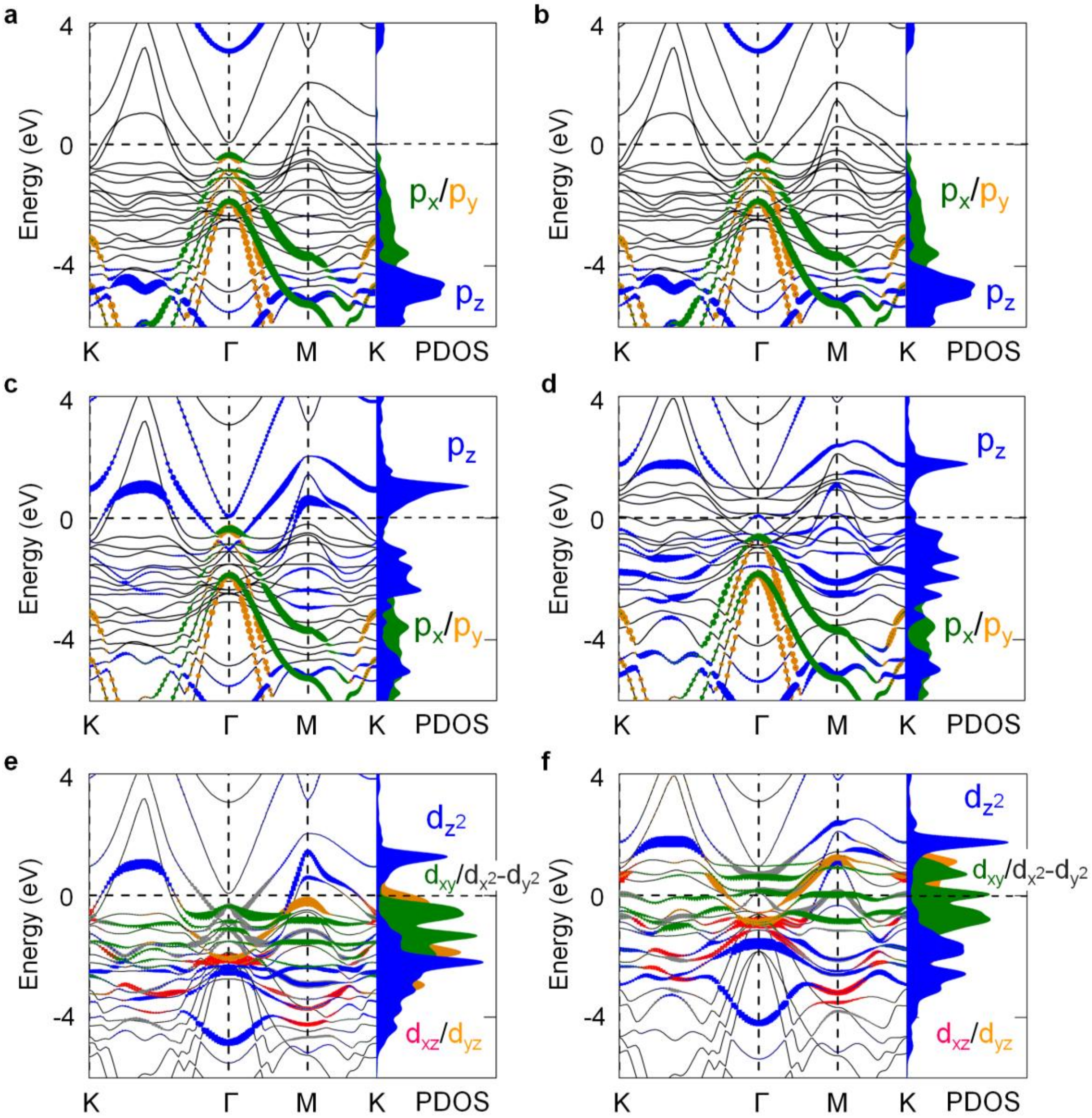

**Figure 4:**

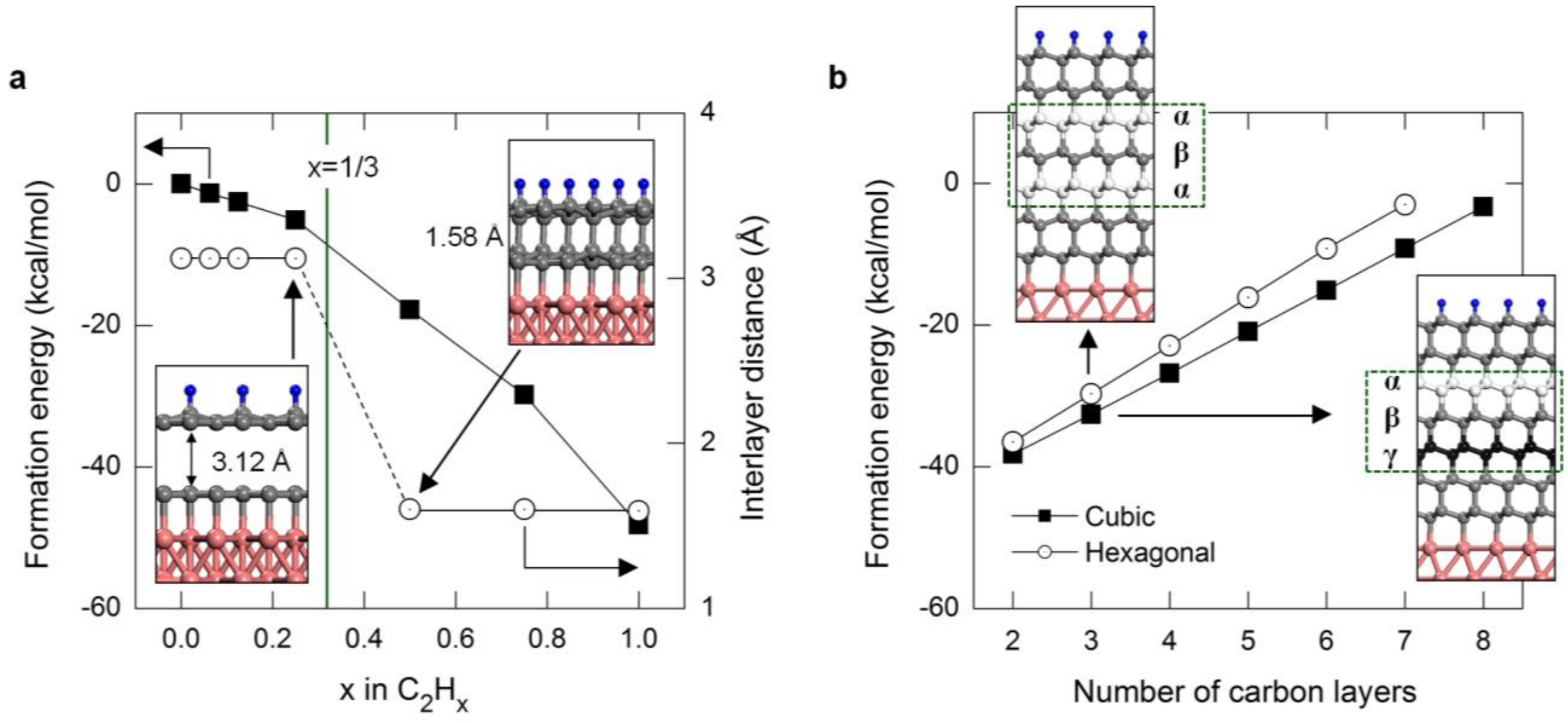

**Figure 5:**

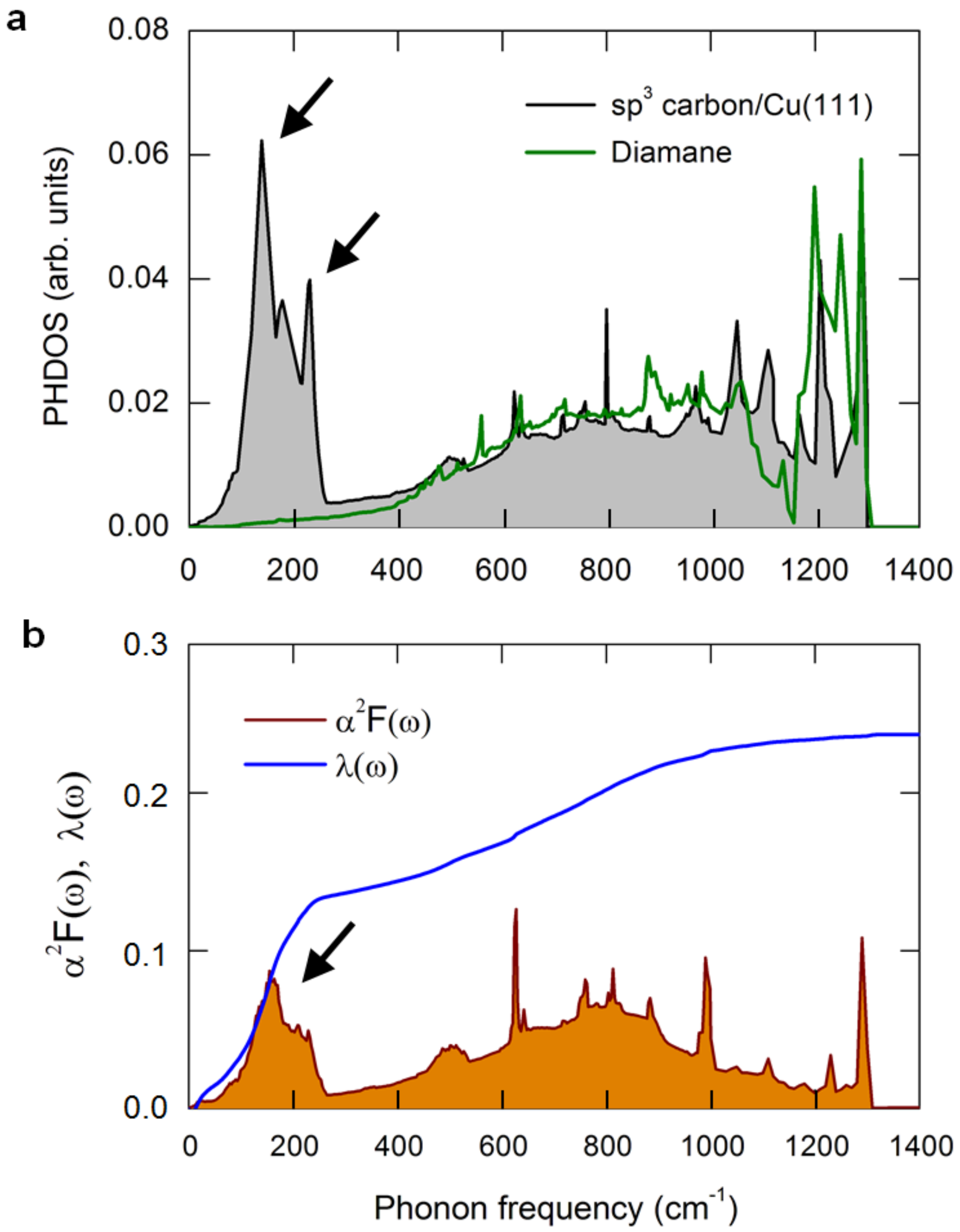